\documentclass[aps,twocolumn, preprintnumbers,amsmath,showpacs,amssymb,prl,nofootinbib]{revtex4-1}

\usepackage{graphicx}
\usepackage{dcolumn}
\usepackage{bm}

\newcommand{\be}{\begin{equation}}
\newcommand{\ee}{\end{equation}}
\newcommand{\ba}{\begin{eqnarray}}
\newcommand{\ea}{\end{eqnarray}}

\begin{document}

\title{Predictions for multiplicities and flow harmonics in 5.44 TeV Xe+Xe collisions at the CERN Large Hadron Collider} 

\author{K.~J.~Eskola${}^{a,b}$, H.~Niemi${}^{a,b,c}$,   R.~Paatelainen${}^{d,b}$, K.~Tuominen${}^{d,b}$}
\affiliation{$^{a}$University of Jyvaskyla, Department of Physics, P.O. Box 35, FI-40014 University of Jyvaskyla, Finland}
\affiliation{$^b$Helsinki Institute of Physics, P.O.Box 64, FI-00014 University of Helsinki, Finland}
\affiliation{$^c$Institut f\"ur Theoretische Physik, Johann Wolfgang Goethe-Universit\"at,
Max-von-Laue-Str.~1, D-60438 Frankfurt am Main, Germany}
\affiliation{$^d$Department of Physics, P.O.Box 64, FI-00014 University of Helsinki, Finland} 
\preprint{HIP-2017-33/TH}

\begin{abstract}
We present the next-to-leading-order event-by-event EKRT model predictions for the centrality dependence of the charged hadron multiplicity in the pseudorapidity interval $|\eta|\le 0.5$, and for the centrality dependence of the charged hadron flow harmonics $v_n\{2\}$ obtained from 2-particle cumulants, in $\sqrt{s_{NN}}=5.44$ TeV Xe+Xe collisions at the CERN Large Hadron Collider. Our prediction for the 0-5 \% central charged multiplicity is $dN_{\rm ch}/d\eta =1218\pm 46$. We also predict $v_n\{2\}$ in Xe+Xe collisions to increase more slowly from central towards peripheral collisions than those in a Pb+Pb system. We find that at $10 \dots 50$\% centralities  $v_2\{2\}$ is smaller and $v_3\{2\}$ is larger than in the Pb+Pb system while  $v_4\{2\}$ is of the same magnitude in both systems.  We also find that the ratio of flow harmonics in Xe+Xe collisions and in Pb+Pb collisions shows a slight sensitivity to the temperature dependence of the shear-viscosity-to-entropy ratio. As we discuss here, the new nuclear mass-number systematics especially in the flow harmonics serves as a welcome further constraint for describing the space-time evolution of a heavy-ion system and for determining the shear viscosity and other transport properties of strongly interacting matter.
\end{abstract}

 
\maketitle 

\section{Introduction}
Ultrarelativistic heavy-ion collisions at the CERN Large Hadron Collider (LHC) and BNL Relativistic Heavy Ion Collider (RHIC) probe the collectivity in Quantum Chromodynamics (QCD) by producing strongly-interacting QCD matter at high temperatures and vanishing net-baryon number densities. To correctly interpret the measurements, it is of pivotal importance to understand the primary production dynamics of the Quark-Gluon Plasma (QGP) and know the transport properties of the produced QCD matter. For this, on the one hand, one needs QCD-based \textit{predictive} modeling to describe the production of the system at various collision energies and nuclei \cite{Eskola:1988yh,McLerran:1993ni,Eskola:1999fc,Kharzeev:2000ph,Kharzeev:2001gp,Lappi:2003bi,Drescher:2006ca,Gelis:2010nm,Albacete:2010ad,Paatelainen:2012at,Paatelainen:2013eea}. Combining this with a fluid-dynamical space-time evolution, event by event, then enables the computation of a multitude of small transverse-momentum ($p_T$) final-state observables \cite{Schenke:2010rr,Gale:2012rq,Schenke:2012wb,Pierog:2013ria,Shen:2014vra,Karpenko:2015xea,Niemi:2015qia,Ryu:2015vwa,Noronha-Hostler:2015uye,Giacalone:2016afq,Gardim:2016nrr,Giacalone:2017dud}. On the other hand, for extracting the \textit{uncertainties} of the QCD matter properties and other theory-input parameters from the measurements, one needs a statistical multi-observable (global) analysis \cite{Novak:2013bqa,Pratt:2015zsa,Bernhard:2016tnd,Bass:2017zyn,Bernhard:2017vql,Auvinen:2017fjw}.

The Bayesian global analysis of small-$p_T$ observables in the LHC Pb+Pb collisions, discussed in Refs.~\cite{Bernhard:2016tnd,Bass:2017zyn,Bernhard:2017vql}, quite strongly suggests that the initial-state dynamics in heavy-ion collisions is correctly captured by the saturation models Next-to-leading-order (NLO) event-by-event (EbyE) EKRT \cite{Niemi:2015qia}\footnote{\protect Named after the authors of Ref.~\cite{Eskola:1999fc}.} and IP-Glasma \cite{Schenke:2012wb}. Interestingly, even though these models approach the saturation from different limits -- EKRT from collinear factorization and perturbative QCD (pQCD), and IP-Glasma from classical gluon fields and small-$x$ QCD -- it is the same phenomenon, dominance of gluon fusions (non-linearities) at small transverse-momenta, that regulates and controls the produced initial multiplicities in both frameworks.

\begin{figure}[!h]
\includegraphics[height=6.20cm]{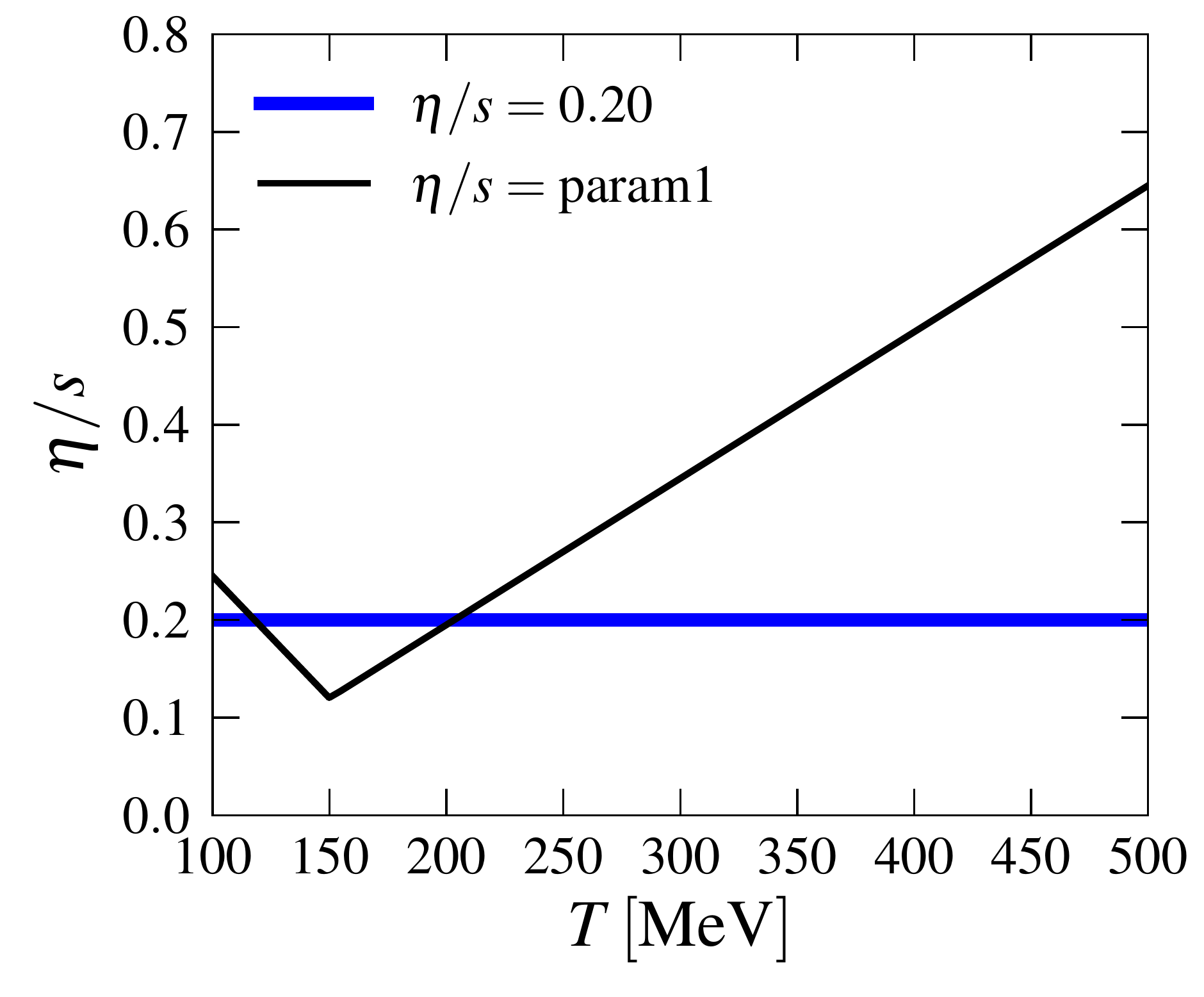}
\caption{\protect The two parametrizations of the temperature dependence of the shear-viscosity-to-entropy ratio which in the NLO EbyE EKRT model \cite{Niemi:2015qia,Niemi:2015voa} give the best overall fit to the various small-$p_T$ LHC and RHIC observables. 
}
\label{fig:etapers}
\end{figure}

The NLO EbyE EKRT model, where the computed QCD-matter initial states are combined with  shear-viscous fluid dynamics, predicts successfully the centrality- and cms-energy dependencies of hadron multiplicities and their $p_T$ distributions, flow coefficients, relative elliptic flow fluctuations, and various flow-correlators in 2.76 and 5.02 TeV Pb+Pb collisions at the LHC and 200 GeV Au+Au collisions at RHIC \cite{Niemi:2015qia,Niemi:2015voa}. For a recent review, see Ref.~\cite{Eskola:2017imo}. Importantly, the simultaneous EKRT analysis of all these observables suggests systematically that the temperature dependence of the shear-viscosity-to-entropy ($\eta/s$) ratio of the produced QCD matter is in the range shown by Fig.~\ref{fig:etapers}. As expected, similar results for the range of $\eta/s(T)$ were then obtained also in the Bayesian analysis -- see Ref.~\cite{Bass:2017zyn} for a recent review. 

What has been missing so far, however, is the nuclear mass-number ($A$) systematics at the LHC. The short $^{129}$Xe+$^{129}$Xe  run with nucleon-nucleon cms-energy $\sqrt{s_{NN}}=5.44$ TeV at the LHC in October 2017 now fills this gap conveniently, offering important further observables to be included in the global analyses. Motivated by this LHC run, in this letter we will present our NLO EbyE EKRT model predictions for the centrality dependencies of the charged hadron multiplicities in the pseudorapidity interval $|\eta|\le 0.5$ and of the flow coefficients $v_n\{2\}$ computed from the 2-particle cumulants. These predictions are obtained by using the two $\eta/s(T)$ parametrizations of Fig~\ref{fig:etapers}. In addition to the $A$ systematics, it will be interesting to see whether the $v_n\{2\}$ in the Xe+Xe collisions would show any increased sensitivity to $\eta/s(T)$ relative to the larger Pb+Pb system. 

Below, we will first outline the procedure for computing these predictions, then show and discuss the obtained results. A detailed account of the model details can be found in Ref.~\cite{Niemi:2015qia}. 

\section{Key steps for Xe+Xe predictions}
\label{sec:keysteps}

\textbf{The first step} is to compute the transverse-area density of minijet transverse energy, $dE_T/d^2\mathbf{r}$, in a nuclear collision  at a given cms-energy $\sqrt{s_{NN}}$ and impact parameter \textbf{b}, accounting for minijets above a transverse momentum cut-off $p_0\gg\Lambda_{\rm QCD}$ in a central rapidity unit $\Delta y$, 
\begin{equation}
\frac{dE_T}{d^2{\bf r}} = T_A(\mathbf{r}_1)T_A(\mathbf{r}_2)
\sigma\langle E_T \rangle_{p_0, \Delta y,\beta}.
\label{eq: dET}
\end{equation}
The nuclear thickness functions $T_A$ give the collision geometry, with $\mathbf{r}_{1,2} = \mathbf{r} \mp \mathbf{b}/2$ where $\mathbf{r}$ is the transverse coordinate. For Xe (isotope $A=129$) we compute the $T_A$ using the Woods-Saxon nuclear density with standard radius parameters $R_A = 1.12 A^{1/3} - 0.86 A^{-1/3}$ and  $d=0.54$ fm.
The $E_T$-weighted minijet cross section $\sigma\langle E_T \rangle_{p_0, \Delta y,\beta}$ \cite{Eskola:1988yh} is computed in collinear factorization and NLO pQCD using the subtraction method \cite{Kunszt:1992tn,Eskola:2000ji,Eskola:2000my}.
This involves the squared invariant amplitudes for $2\rightarrow 2$  and $2\rightarrow 3$ parton scatterings \cite{Ellis:1985er,Paatelainen:2014fsa}, CTEQ6M parton distribution functions \cite{Pumplin:2002vw} supplemented with EPS09s transverse-coordinate dependent nuclear effects \cite{Helenius:2012wd}. Also included in $\sigma\langle E_T \rangle$ are the infrared(IR)- and collinear(CL)-safe measurement functions which define the minijet $E_T$ as the scalar sum of minijet transverse momenta in $\Delta y$, as well as IR/CL-safe definitions of the cut-off $p_0$ and of the minimum $E_T$ ($E_T\ge \beta p_0$; $0\le \beta\le 1$) allowed in $\Delta y$ \cite{Eskola:2000ji,Eskola:2000my,Paatelainen:2012at}. 
For a detailed formulation of $\sigma\langle E_T \rangle$, see Refs.~\cite{Paatelainen:2012at,Niemi:2015qia}. 
In the first step, we thus compute the $\sigma\langle E_T \rangle$ various times: for an array of $\beta$, for a range of $p_0$ values around the expected saturation momentum, for various values of $\mathbf{b}$ and for a lattice of $\mathbf{r}$ in the first quadrant of the transverse plane.

\begin{figure}[!h]
\includegraphics[width=8.20cm,height=6.20cm]{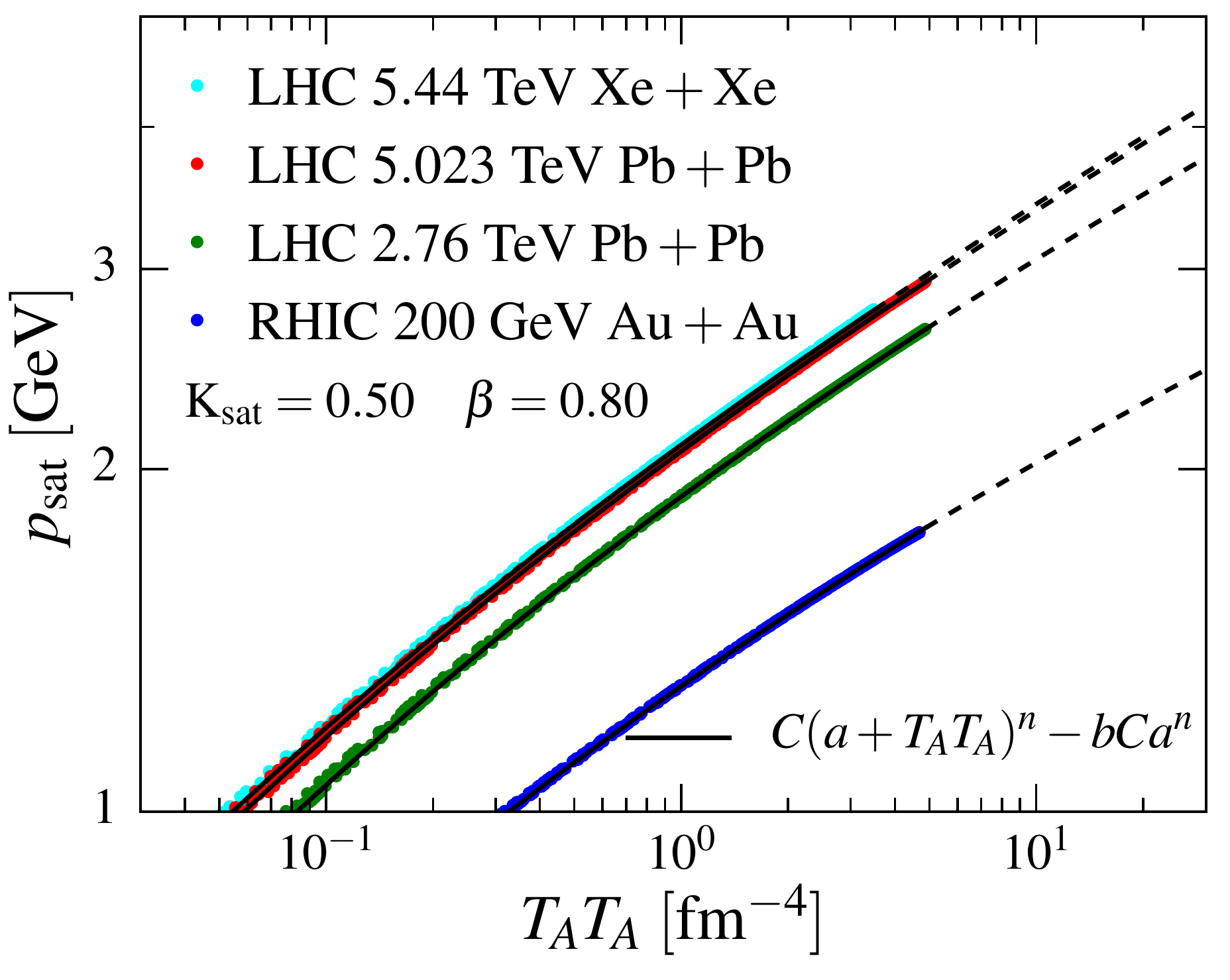}
\caption{\protect Saturation momenta vs. the thickness function product, computed in 
 5.44 TeV Xe+Xe collisions for $|\mathbf{b}|$ = 0, 6.59, and 8.27 fm, with $\beta=0.8$ and  $K_{\rm{sat}}=0.5$ corresponding to $\eta/s(T)$ of \textit{param1}.
The previous results \cite{Niemi:2015qia,Niemi:2015voa} for 5.023 and 2.76 TeV Pb+Pb collisions at the LHC, and 200 GeV Au+Au collisions at RHIC are shown for comparison.
The parametrization $p_{\rm sat}(T_AT_A;K_{\rm{sat}},\beta)$ of the Xe+Xe result is from Table \ref{tab:psat_parametrization_5440}, and those for the Pb+Pb and Au+Au cases are from Refs.~\cite{Niemi:2015qia,Niemi:2015voa}.
The order of the results from top to bottom is the same as in the legend. The QCD calculations are presented by points and the parametrizations by solid and dashed lines.}
\label{fig:fitsaturation}
\end{figure}

\textbf{The second step} is then to find the mapping between the collision geometry given by $T_A(\mathbf{r}_1)T_A(\mathbf{r}_2)$ and the saturation scale $p_0 = p_{\rm sat}(\sqrt{s_{NN}},A,\mathbf{r},\mathbf{b};\beta,K_{\rm sat})$ 
 which dictates gluon production locally at each $\mathbf{r}$. The saturation momenta are obtained as the solutions of the following saturation criterion \cite{Paatelainen:2012at,Paatelainen:2013eea} which derives from the limit where the minijet $E_T$ production starts to be dominated by higher-order fusion processes $(n\ge 2)\rightarrow 2$ over the usual $2\rightarrow 2$ ones:
\begin{equation}
\label{eq: saturation}
\frac{{\rm d}E_T}{{\rm d}^2{\bf r}}(p_0,\sqrt{s_{NN}},A, \Delta y,{\bf r},{\bf b},\beta) = \frac{K_{\rm{sat}}}{\pi} p_0^3 \Delta y.
\end{equation}
Here $K_{\rm{sat}}$ is a proportionality constant whose value we now know from Ref.~\cite{Niemi:2015qia} where it was obtained by normalizing the computed  charged-particle multiplicity $dN_{\rm ch}/d\eta$ to the one measured by ALICE in $|\eta|\le 0.5$ in 0-5 \% central 2.76 TeV Pb+Pb collisions. Figure \ref{fig:fitsaturation} shows the values of $p_{\rm sat}$ obtained for the current LHC 5.44 TeV Xe+Xe case for three different values of $|\mathbf{b}|$ with $K_{\rm{sat}} = 0.5$ and $\beta = 0.8$ that correspond to our $\eta/s=param1$ parametrization.  Our earlier results for the LHC 5.023 TeV and 2.76 TeV Pb+Pb collisions and RHIC 200 GeV Au+Au collisions from Refs.~\cite{Niemi:2015qia,Niemi:2015voa} are shown for comparison. As expected \cite{Eskola:2001rx,Niemi:2015qia}, $p_{\rm sat}$ in the Xe+Xe system depends again only on the product $T_A(\mathbf{r}_1)T_A(\mathbf{r}_2)$ and not on $\mathbf{r}$ and $\mathbf{b}$ separately, i.e. $p_{\rm sat}(\mathbf{r},\mathbf{b})\approx p_{\rm sat}(T_A(\mathbf{r}_1)T_A(\mathbf{r}_2))$. This is the key feature making our EbyE framework possible in practice.

For handling the densest nuclear overlap regions in the fluctuating EbyE case (see Fig.~\ref{fig:fitsaturation}, dashed lines), we must still parametrize the $T_AT_A$ dependence of $p_{\rm sat}$. For enabling further studies of the uncertainties related to the fixing of $K_{\rm{sat}}$ and $\beta$,  we also parametrize the $K_{\rm{sat}}$ and $\beta$ dependence of $p_{\rm sat}$ similarly to what we did before \cite{Niemi:2015qia,Niemi:2015voa}. The outcome is shown in Table \ref{tab:psat_parametrization_5440} and Fig.~\ref{fig:fitsaturation}. 

\begin{table}[h]
\caption{Parametrization  
$p_{\rm sat}(T_AT_A; K_{\rm sat}, \beta) = C\left[a + \rho_{AA}\right]^n - b C a^n$
in 5.44 TeV Xe+Xe collisions for $K_{\rm sat} \in [0.4, 2.0]$ and $\beta<0.9$.
The ($K_{\rm sat}, \beta$) dependence of $a$, $b$, $C$ and $n$ (see below) is parametrized as 
$P_i(K_{\rm sat}, \beta) = a_{i0} + a_{i1} K_{\rm sat} + a_{i2} \beta + a_{i3}K_{\rm sat}\beta + a_{i4}\beta^2 + a_{i5} K_{\rm sat}^2$. }
\begin{tabular}{c|cccc}
\hline
\hline
$P_i\rightarrow$ & $C$          & $n$                  & $a$             & $b$ \\  
\hline
$a_{i0}$	& 3.9593855    & 0.1454605      & -0.0034895    & 0.8309836 \\
$a_{i1}$	& -0.7184810   & -0.0176172     & 0.0162876     & -0.0592489 \\
$a_{i2}$	& 0.6052426    & -0.0244353     & -0.0004785    & 0.0869927 \\
$a_{i3}$	& 0.0748967    & -0.0030068     & 0.0074064     & -0.0000669 \\
$a_{i4}$	& -1.5647924   & 0.0546230      & -0.0023656    & -0.2067018 \\
$a_{i5}$	& 0.1347936    & 0.0057852      & -0.0026187    & 0.0234235 \\
\hline
\hline
\end{tabular}
\label{tab:psat_parametrization_5440}
\end{table}

\textbf{The third step} is to prepare the initial states for fluid dynamics event by event. As in Refs.~\cite{Niemi:2015qia,Niemi:2015voa}, we obtain the EbyE fluctuating initial energy densities by setting a Gaussian gluon transverse density distribution of a width $\sigma=0.43$ fm around each nucleon sampled from the Woods-Saxon density distribution. The thickness functions $T_A$ are then computed by locally summing up the gluon transverse densities. For fixed $K_{\rm sat}, \beta$, the obtained product $T_AT_A$ now maps to $p_{\rm sat}$  according to Fig.~\ref{fig:fitsaturation} and Table \ref{tab:psat_parametrization_5440}, and we can compute the local energy density at the local formation time $\tau_s(\mathbf{r})=1/p_{\rm sat}(\mathbf{r})$ as
\begin{equation}
\varepsilon(\mathbf{r}, \tau_s(\mathbf{r}) ) = \frac{dE_T(p_{\rm sat})}{d^2{\bf r}} \frac{1}{\tau_s(\mathbf{r})\Delta y} = \frac{K_{\rm{sat}}}{\pi}[p_{\rm sat}(\mathbf{r})]^4.
\label{eq:edensity}
\end{equation}
We assume this to be valid for $p_{\rm sat}\ge p_{\rm sat}^{\rm min}= 1$ GeV. To start the fluid-dynamic simulation at a constant proper time $\tau_0=1/p_{\rm sat}^{\rm min}=0.2$ fm, we evolve these energy densities locally from $\tau_s(\mathbf{r})$ to 0.2 fm with 1 D Bjorken hydrodynamics.  Finally, we treat the dilute edges of the system as explained in \cite{Niemi:2015qia} (Eq. (34) there\footnote{\protect Now $\sigma_{NN}=70.53$ mb, obtained from the parametrization in Refs.~\cite{Cudell:2002xe,Antchev:2013iaa}.}).

\textbf{The fourth step} is the fluid-dynamic runs, event by event. Our setup is identical to that of Refs.~\cite{Niemi:2015qia,Niemi:2015voa}, i.e. we apply 2nd-order dissipative relativistic 2+1 D hydro with transient fluid-dynamics equation of motion for the shear-stress tensor as in Refs. \cite{Denicol:2012cn,Molnar:2013lta}. Our QCD-matter equation of state is \textit{s95p}-PCE-v1 \cite{Huovinen:2009yb} with chemical decoupling taking place at $T_{\rm chem} = 175$ MeV, and kinetic freeze-out at $T_{\rm dec} = 100$ MeV. Initially the shear-stress tensor $\pi^{\mu\nu}$ and transverse flow are assumed to be zero, and on the freeze-out surface the viscous $\delta f$ corrections are $\propto p_\mu p_\nu \pi^{\mu\nu}$. We do not consider the bulk viscosity or heat conductivity but account for the shear viscosity and its temperature dependence as shown in Fig.~\ref{fig:etapers}.

\section{Results}
\label{sec:results}

\begin{figure}[b!h]
\hspace{-.8cm} 
\includegraphics[width=8.60cm]{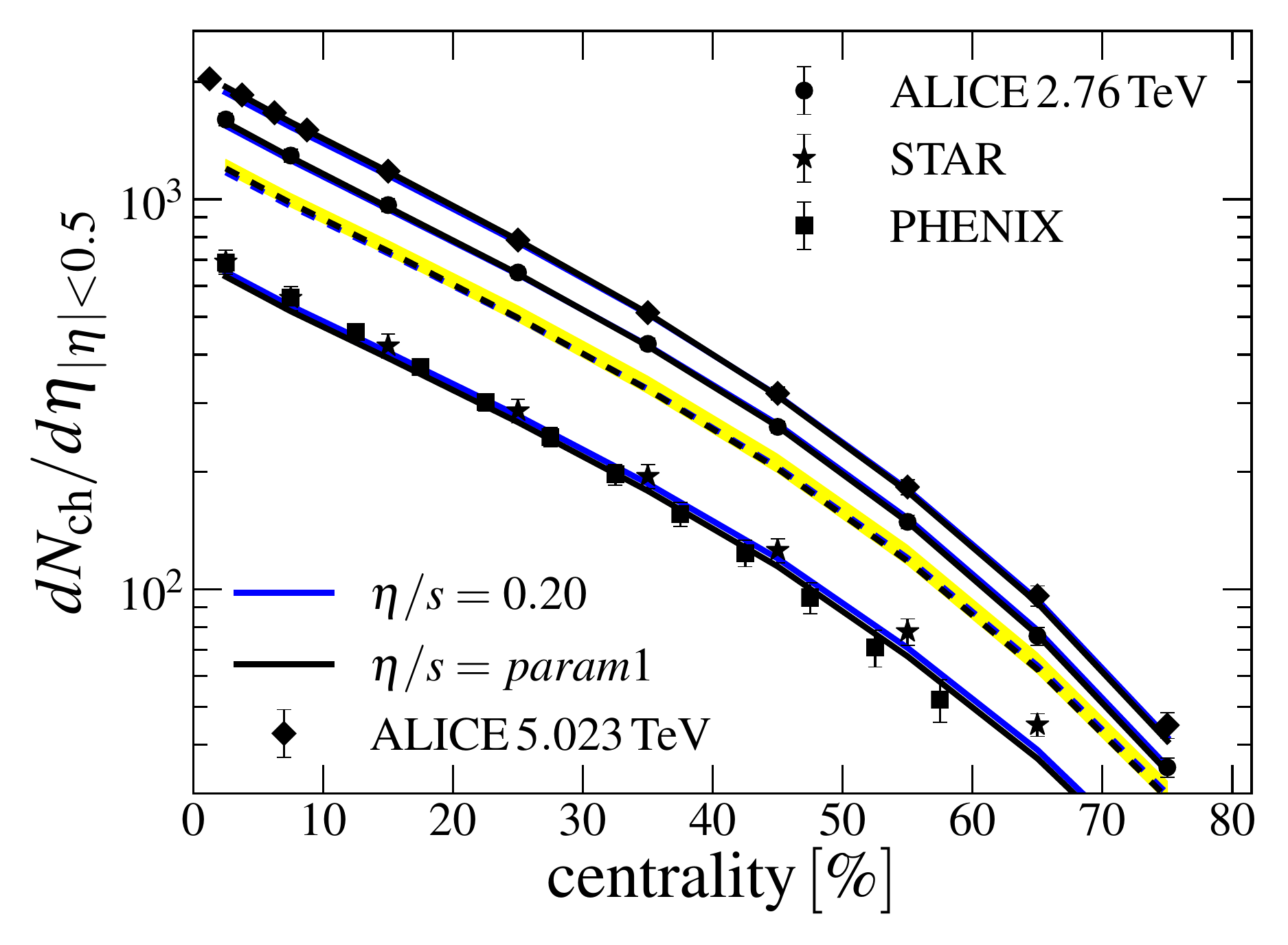}
\caption{\protect The NLO EbyE EKRT model prediction for the centrality dependence of charged hadron multiplicity in 5.44 TeV Xe+Xe collisions at the LHC, computed for the two $\eta/s(T)$ parametrizations of Fig.~\ref{fig:etapers}.  
The results of Refs.~\cite{Niemi:2015qia,Niemi:2015voa}, the Pb+Pb measurements by ALICE at 2.76 and 5.02 TeV \cite{Aamodt:2010cz,Adam:2015ptt}, and the Au+Au measurements at 200 GeV by STAR \cite{Abelev:2008ab} and PHENIX \cite{Adler:2004zn} are also shown. 
The computed centralmost multiplicities for 5.44 (2.76) TeV are 1172 (1542) for $\eta/s=0.2$, and 1199 (1576) for \textit{param1}, while the ALICE 2.76 TeV measurement gives $1601\pm60$ \cite{Aamodt:2010cz}.
The error band on our 5.44 TeV prediction accounts for the normalization uncertainty as explained in the text.  
}
\label{fig:multiplicity}
\end{figure}

Figure~\ref{fig:multiplicity} shows our prediction for the centrality dependence of the charged particle multiplicity in 5.44 TeV Xe+Xe collisions, computed for the $\eta/s$ parametrizations of Fig.~\ref{fig:etapers}.
Our previous results for 2.76 TeV Pb+Pb collisions and  200 GeV Au+Au collisions \cite{Niemi:2015qia}, and predictions for 5.023 TeV Pb+Pb collisions \cite{Niemi:2015voa}, as well as the ALICE data \cite{Aamodt:2010cz,Adam:2015ptt} for 2.76 and 5.02 TeV, and the STAR \cite{Abelev:2008ab} and PHENIX \cite{Adler:2004zn} data for the 200 GeV collisions, are shown for comparison. 

We emphasize that the centralmost ALICE datapoint at 2.76 TeV -- and this one point only -- has been used for fixing the normalization of our results ($K_{\rm sat}= 0.63\, (0.5)$ for $\eta/s=0.2$ (\textit{param1}) from \cite{Niemi:2015qia}). Thus, the error bar of this ALICE point translates into a normalization uncertainty in our 5.44 TeV prediction as shown by the yellow error band. The normalization was done iteratively to a few percent accuracy, causing the corresponding difference between the predicted centralmost 5.44 TeV multiplicities (dashed curves). We refrain here from further finetuning, because the relative normalization uncertainties in our centralmost 2.76 TeV multiplicity are transmitted practically directly into those of our 5.44 TeV multiplicity prediction. We account for these normalization uncertainties as follows,
using the centralmost multiplicities given in the caption of Fig.~\ref{fig:multiplicity}:  Multiplying the \textit{param1} result 1199 by 1601/1576 gives 1218, and estimating the error as $(60/1601)\times1218 = 46$,
we arrive at our best prediction for the charged multiplicity in 0-5 \% central 5.44 TeV Xe+Xe collisions: 
\begin{equation}
\frac{dN_{\rm ch}}{d\eta}\bigg|_{|\eta|\le 0.5}^{\rm Xe+Xe} =1218\pm 46.
\end{equation} 
For the $\eta/s=0.2$ case the result is essentially the same. The error band on our 5.44 TeV prediction in Fig.~\ref{fig:multiplicity} is obtained with this procedure, applying the same relative normalization uncertainty for all centralities and individually for both $\eta/s$ parametrizations.

\begin{figure}[!h]
\hspace{-0.5cm} 
\includegraphics[width=8.20cm]{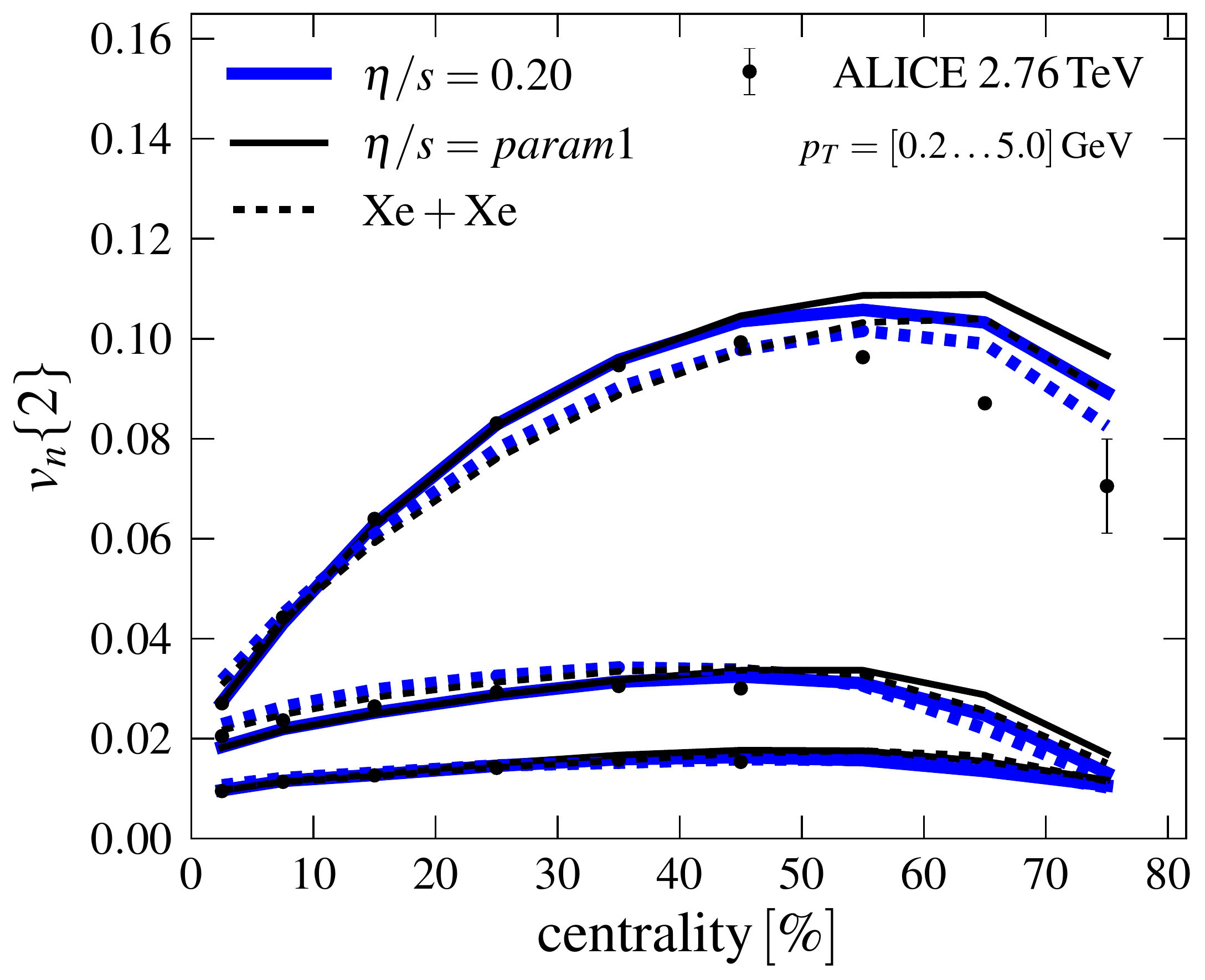}
\caption{\protect The NLO EbyE EKRT model prediction for the centrality dependence of charged hadron 
flow harmonics $v_n\{2\}$  in 5.44 TeV Xe+Xe collisions at the LHC, computed for the $\eta/s(T)$ parametrizations of Fig.~\ref{fig:etapers}. The results from Ref.~\cite{Niemi:2015qia} and ALICE data \cite{ALICE:2011ab} for 2.76 TeV Pb+Pb collisions are also shown. 
}
\label{fig:vn_cent}
\end{figure}
\begin{figure*}[!]
\hspace{-0.5cm} 
\includegraphics[width=16.20cm]{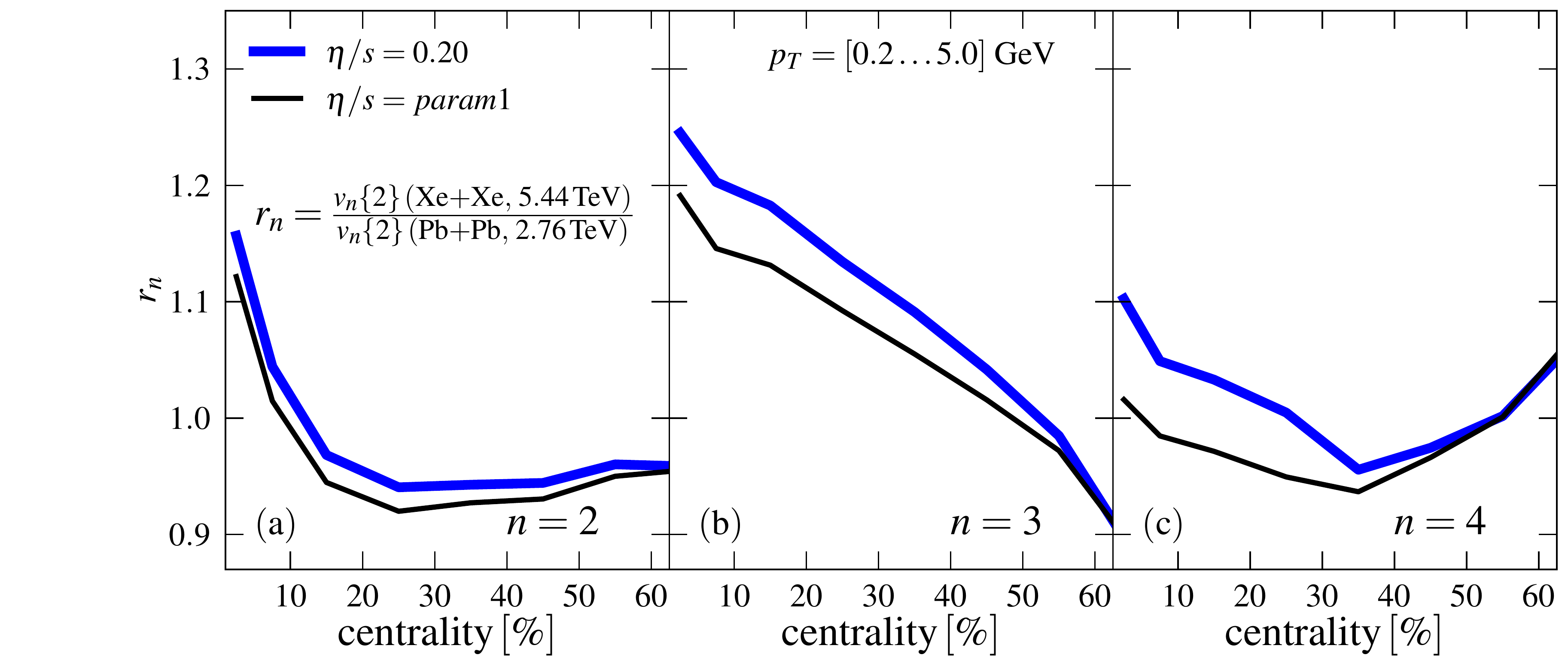}
\caption{\protect 
Ratios of the predicted flow harmonics $v_n\{2\}$  in  5.44 TeV Xe+Xe and 2.76 Pb+Pb collisions, for each of the $\eta/s$ parametrizations studied. 
}
\label{fig:vn_cent_ratio}
\end{figure*}

Our 5.44 TeV Xe+Xe prediction for the centrality dependence of the 2-particle cumulant flow harmonics $v_n\{2\}$ of charged hadrons is shown in Fig.~\ref{fig:vn_cent} along with our earlier results for 2.76 TeV Pb+Pb collisions \cite{Niemi:2015qia} and the corresponding ALICE data \cite{ALICE:2011ab}.
Perhaps not readily expected, the $v_2\{2\}$ in Xe+Xe collisions is predicted to increase more slowly towards peripheral collisions than that in the Pb+Pb system. We also predict that from central to semi-peripheral collisions  $v_3\{2\}$ is larger in Xe+Xe than in Pb+Pb, which could be expected since the $v_3\{2\}$ originates from the initial density fluctuations of the system which are larger in a smaller system. Finally, we predict $v_
4\{2\}$ to be of the same magnitude in both systems but show a flattening of the centrality slope in Xe+Xe. To quantify the  systematics of the predicted flow-coefficients as well as the sensitivity to $\eta/s(T)$ in more detail, we also plot the corresponding ratios  $v_n\{2\}(\text{Xe+Xe, 5.44 TeV}) / v_n\{2\}(\text{Pb+Pb,  2.76 TeV})$ in Fig.~\ref{fig:vn_cent_ratio}. We can see from Figs.~\ref{fig:vn_cent} and \ref{fig:vn_cent_ratio} that a simultaneous analysis of the flow harmonics in Xe+Xe and Pb+Pb collisions shows more sensitivity to $\eta/s(T)$ than the analysis of Pb+Pb collisions alone.

\begin{figure}[!h]
\hspace{-0.5cm} 
\includegraphics[width=8.20cm]{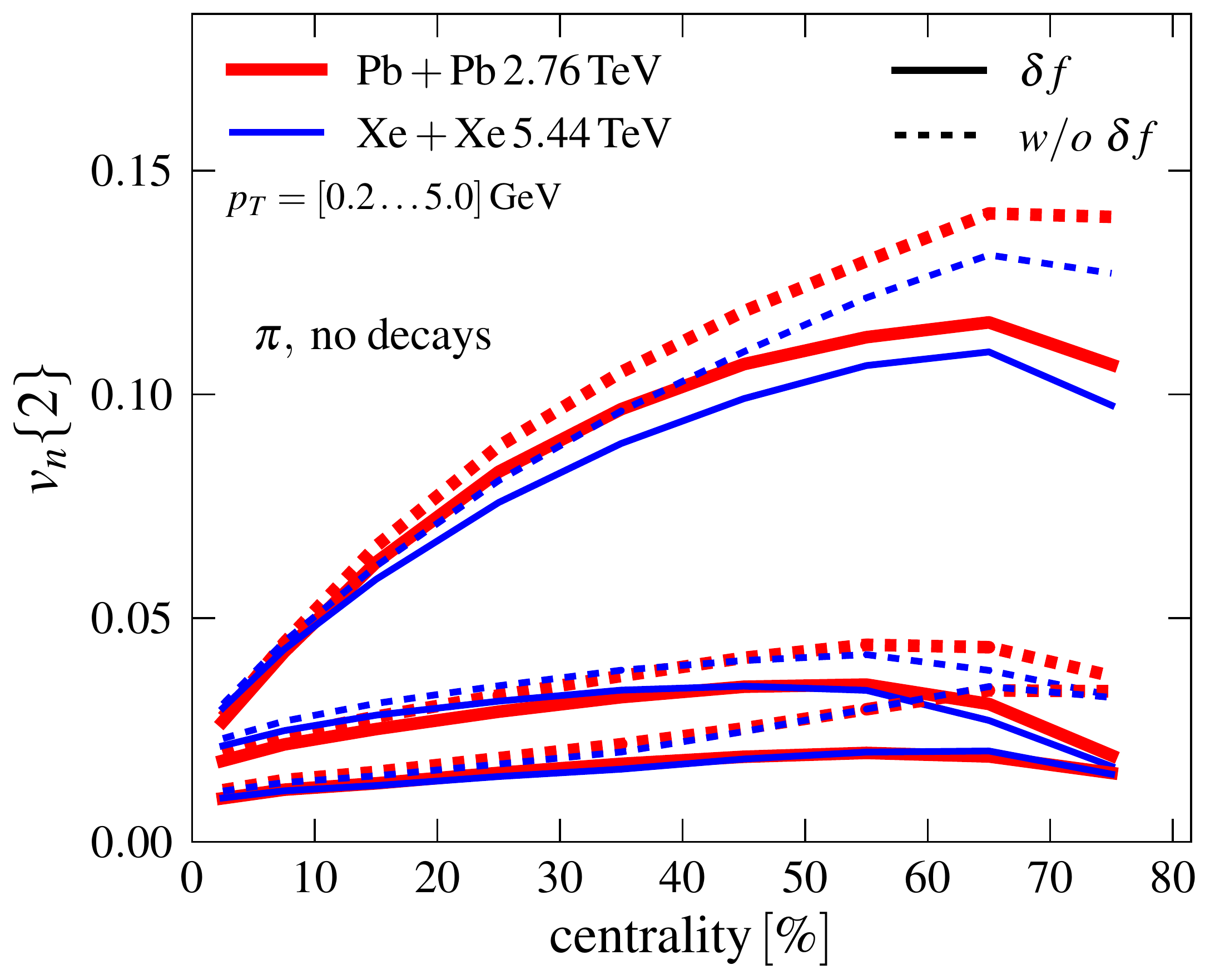}
\caption{\protect The effect of the $\delta f$ corrections in the 2-particle cumulant flow harmonics in  
5.44 TeV Xe+Xe (thin blue curves) and 2.76 TeV Pb+Pb collisions (thick red curves) vs. centrality. Only thermal pions (no pions from decays) are considered in this figure.}
\label{fig:deltaf}
\end{figure}

To analyze the predicted $v_n\{2\}$ systematics further, we study in Fig.~\ref{fig:deltaf} the effects of the $\delta f$ corrections for the Xe+Xe and Pb+Pb systems. We see that these effects are similar in the relative magnitude and that they remain acceptably small until $\sim$50 \% centralities in both systems. Thus, we can confirm that the predicted slope change of $v_n\{2\}$ (and of $v_2\{2\}$ in particular) from Pb+Pb to Xe+Xe is indeed \textit{not} caused by the $\delta f$ corrections. 
%

\begin{figure}[!]
\hspace{-0.5cm} 
\includegraphics[width=7.20cm]{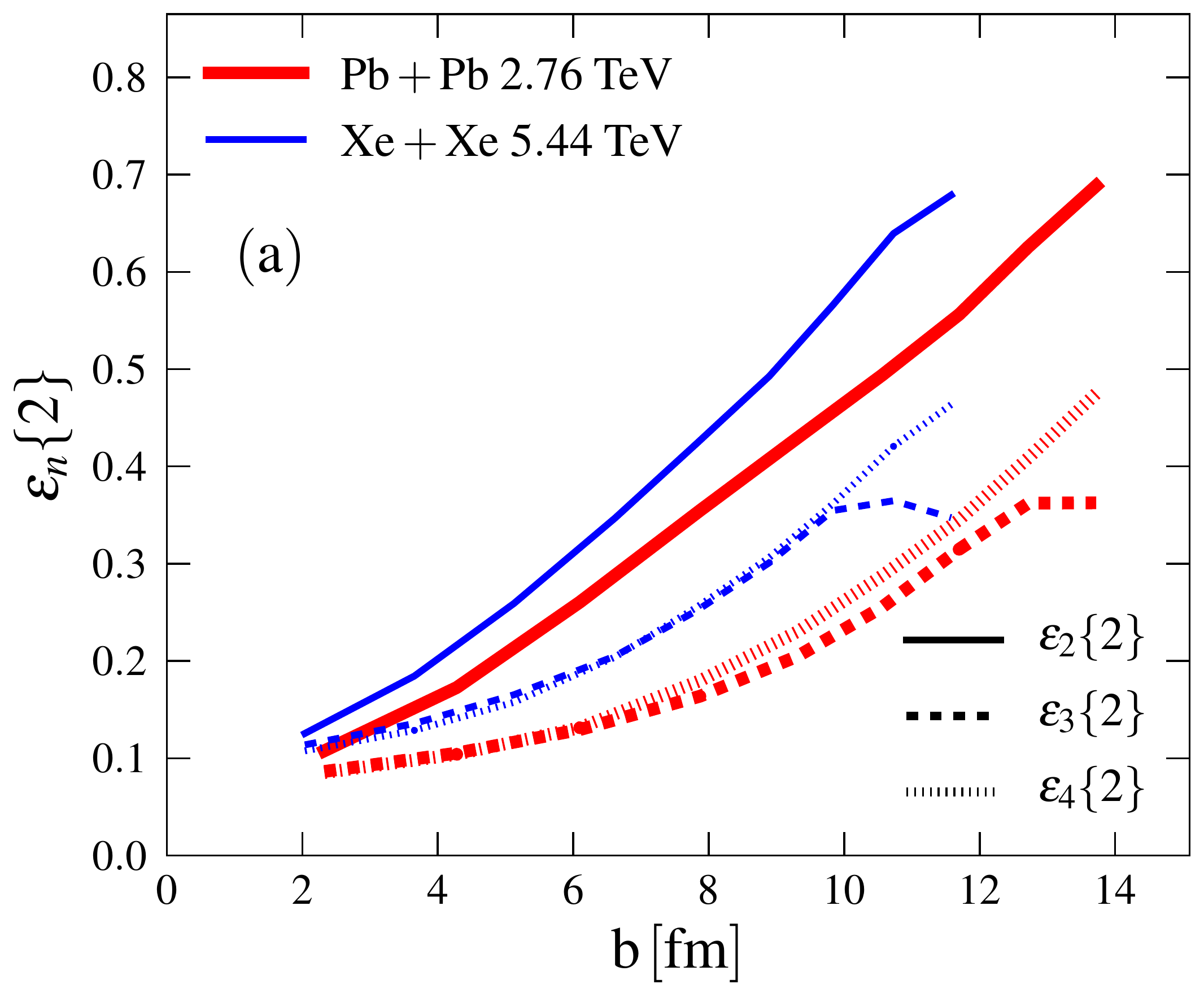}
\includegraphics[width=7.20cm]{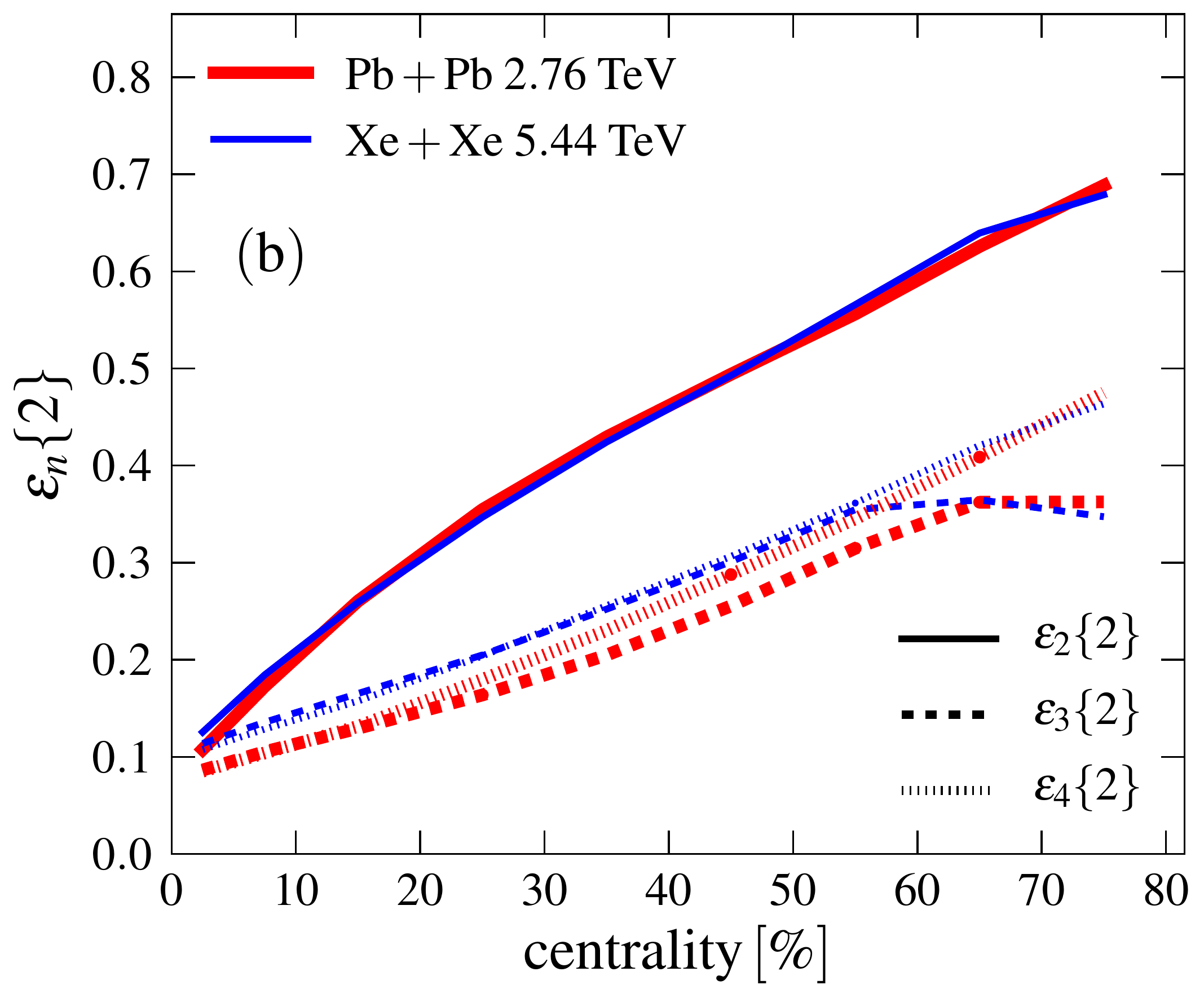} 
\caption{\protect Initial eccentricities $\varepsilon_n\{2\}$ in 5.44 TeV Xe+Xe (thin blue curves) and 2.76 TeV Pb+Pb collisions (thick red curves) vs. impact parameter (upper panel) and vs. centrality (lower panel). 
}
\label{fig:eps_n}
\end{figure}

Figure \ref{fig:eps_n} then shows the initial eccentricities $\varepsilon_n\{2\}$ as a function of the impact parameter and centrality. As expected, $\varepsilon_n\{2\}$ against $|\mathbf{b}|$ shows larger eccentricities for the smaller system. However, when $\varepsilon_n\{2\}$ is considered against centrality, this systematics changes. This is because the average impact parameter in each centrality class is smaller for a smaller system. This relative shift in $\langle|\mathbf{b}|\rangle$ when moving from Pb+Pb to Xe+Xe collisions, together with the $|\mathbf{b}|$-slope systematics of $\varepsilon_n\{2\}$, explains why $\varepsilon_2\{2\}$ vs. centrality becomes of the same magnitude in both systems while $\varepsilon_{3,4}\{2\}$ remain slightly larger in the Xe+Xe system. The fluid-dynamical evolution is then responsible for the rest, converting the initial $\varepsilon_n\{2\}$ into the final-state $v_n\{2\}$ the more efficiently the longer the system evolves in the QGP phase where the pressure gradients are large. Thus, towards peripheral Xe+Xe collisions, 
and with larger $\eta/s$ in the QGP, the $\varepsilon_n\{2\}\rightarrow v_n\{2\}$ conversion efficiency decreases. This explains why the $v_n\{2\}$ slopes from central towards peripheral collisions are smaller in the Xe+Xe system than in the Pb+Pb system, and why the $v_n\{2\}$ for $\eta/s=0.2$ are larger than those for \textit{param1}.

\section{Conclusions}

We have computed the centrality dependencies of the charged hadron multiplicity and 2-particle cumulant flow harmonics in 5.44 TeV Xe+Xe collisions at the LHC by applying the machinery and predictive power of the NLO EbyE EKRT model, and using our two best-fitting parametrizations of $\eta/s(T)$ of Ref.~\cite{Niemi:2015qia}. 
The corresponding multiplicity now (October 2017) measured at the LHC will serve as a most welcome first constraint for the $A$ systematics of the primary QGP production mechanism at highest collision energies so far. The measurement of flow harmonics in Xe+Xe collisions at 5.44 TeV will in turn offer further constraints for the space-time evolution and $\eta/s(T)$ of QCD matter. The predicted centrality dependence of the flow harmonics in Xe+Xe collisions shows interesting systematics: The centrality slopes of $v_n\{2\}$ are in general smaller in Xe+Xe than in Pb+Pb. From 10 \% central collisions onwards the predicted $v_2\{2\}$ in Xe+Xe becomes smaller than in Pb+Pb collisions, while $v_3\{2\}$ remains larger than in Pb+Pb until 50 \% centralities and $v_4\{2\}$ is of the same magnitude in both systems. 
We look forward to seeing whether the measurements confirm these predictions, especially the Pb+Pb $\rightarrow $ Xe+Xe systematics of the flow-harmonics, which we believe is due to a decreasing efficiency in converting the initial spatial asymmetries (eccentricities) into the final state momentum asymmetries (flow coefficients).

\textit{Acknowledgements}

K.J.E.\ and K.T.\ are supported by the Academy of Finland, Projects 297058 and 310130, correspondingly. H.N.\ is supported by the EU’s Horizon
2020 research and innovation programme under the Marie Sklodowska-Curie grant agreement no.\ 655285, and by the Collaborative Research Center 
CRC-TR 211 “Strong-interaction matter under extreme conditions” funded by DFG. R.P.\ is supported by the European Research Council, grant no.\ 725369. 
We acknowledge the CSC -- IT Center for Science in Espoo, Finland, for the allocation of the computational resources. We also thank the members of 
the ALICE collaboration who drew our attention to the new Xe+Xe run, which motivated this study.


\begin{thebibliography}{50} 

\bibitem{Eskola:1988yh} 
  K.~J.~Eskola, K.~Kajantie and J.~Lindfors,
  Nucl.\ Phys.\ B {\bf 323}, 37 (1989).


\bibitem{McLerran:1993ni} 
  L.~D.~McLerran and R.~Venugopalan,
  Phys.\ Rev.\ D {\bf 49}, 2233 (1994)
  [hep-ph/9309289].


\bibitem{Eskola:1999fc} 
  K.~J.~Eskola, K.~Kajantie, P.~V.~Ruuskanen and K.~Tuominen,
  Nucl.\ Phys.\ B {\bf 570}, 379 (2000)
  [hep-ph/9909456].


\bibitem{Kharzeev:2000ph} 
  D.~Kharzeev and M.~Nardi,
  Phys.\ Lett.\ B {\bf 507}, 121 (2001)
  [nucl-th/0012025].


\bibitem{Kharzeev:2001gp} 
  D.~Kharzeev and E.~Levin,
  Phys.\ Lett.\ B {\bf 523}, 79 (2001)
  [nucl-th/0108006].


\bibitem{Lappi:2003bi} 
  T.~Lappi,
  Phys.\ Rev.\ C {\bf 67}, 054903 (2003)
  [hep-ph/0303076].


\bibitem{Drescher:2006ca} 
  H.-J.~Drescher and Y.~Nara,
  Phys.\ Rev.\ C {\bf 75}, 034905 (2007)
  [nucl-th/0611017].


\bibitem{Gelis:2010nm} 
  F.~Gelis, E.~Iancu, J.~Jalilian-Marian and R.~Venugopalan,
  Ann.\ Rev.\ Nucl.\ Part.\ Sci.\  {\bf 60}, 463 (2010)
  [arXiv:1002.0333 [hep-ph]].

\bibitem{Albacete:2010ad} 
  J.~L.~Albacete and A.~Dumitru,
  arXiv:1011.5161 [hep-ph].

\bibitem{Paatelainen:2012at} 
  R.~Paatelainen, K.~J.~Eskola, H.~Holopainen and K.~Tuominen,
  Phys.\ Rev.\ C {\bf 87}, no. 4, 044904 (2013)
  [arXiv:1211.0461 [hep-ph]].


\bibitem{Paatelainen:2013eea} 
  R.~Paatelainen, K.~J.~Eskola, H.~Niemi and K.~Tuominen,
  Phys.\ Lett.\ B {\bf 731}, 126 (2014)
  [arXiv:1310.3105 [hep-ph]].


\bibitem{Schenke:2010rr} 
  B.~Schenke, S.~Jeon and C.~Gale,
  Phys.\ Rev.\ Lett.\  {\bf 106}, 042301 (2011)
  [arXiv:1009.3244 [hep-ph]].


\bibitem{Gale:2012rq} 
  C.~Gale, S.~Jeon, B.~Schenke, P.~Tribedy and R.~Venugopalan,
  Phys.\ Rev.\ Lett.\  {\bf 110}, no. 1, 012302 (2013)
  [arXiv:1209.6330 [nucl-th]].


\bibitem{Schenke:2012wb} 
  B.~Schenke, P.~Tribedy and R.~Venugopalan,
  Phys.\ Rev.\ Lett.\  {\bf 108}, 252301 (2012)
  [arXiv:1202.6646 [nucl-th]].


\bibitem{Pierog:2013ria} 
  T.~Pierog, I.~Karpenko, J.~M.~Katzy, E.~Yatsenko and K.~Werner,
  Phys.\ Rev.\ C {\bf 92}, no. 3, 034906 (2015)
  [arXiv:1306.0121 [hep-ph]].


\bibitem{Shen:2014vra} 
  C.~Shen, Z.~Qiu, H.~Song, J.~Bernhard, S.~Bass and U.~Heinz,
  Comput.\ Phys.\ Commun.\  {\bf 199}, 61 (2016)
  [arXiv:1409.8164 [nucl-th]].

\bibitem{Karpenko:2015xea} 
  I.~A.~Karpenko, P.~Huovinen, H.~Petersen and M.~Bleicher,
  Phys.\ Rev.\ C {\bf 91}, no. 6, 064901 (2015)
  [arXiv:1502.01978 [nucl-th]].
  
\bibitem{Niemi:2015qia} 
  H.~Niemi, K.~J.~Eskola and R.~Paatelainen,
  Phys.\ Rev.\ C {\bf 93}, no. 2, 024907 (2016)
  [arXiv:1505.02677 [hep-ph]].


\bibitem{Ryu:2015vwa} 
  S.~Ryu, J.-F.~Paquet, C.~Shen, G.~S.~Denicol, B.~Schenke, S.~Jeon and C.~Gale,
  Phys.\ Rev.\ Lett.\  {\bf 115}, no. 13, 132301 (2015)
  [arXiv:1502.01675 [nucl-th]].


\bibitem{Noronha-Hostler:2015uye} 
  J.~Noronha-Hostler, M.~Luzum and J.~Y.~Ollitrault,
  Phys.\ Rev.\ C {\bf 93}, no. 3, 034912 (2016)
  [arXiv:1511.06289 [nucl-th]].


\bibitem{Giacalone:2016afq} 
  G.~Giacalone, L.~Yan, J.~Noronha-Hostler and J.~Y.~Ollitrault,
  Phys.\ Rev.\ C {\bf 94}, no. 1, 014906 (2016)
  [arXiv:1605.08303 [nucl-th]].


\bibitem{Gardim:2016nrr} 
  F.~G.~Gardim, F.~Grassi, M.~Luzum and J.~Noronha-Hostler,
  Phys.\ Rev.\ C {\bf 95}, no. 3, 034901 (2017)
  [arXiv:1608.02982 [nucl-th]].

\bibitem{Giacalone:2017dud} 
  G.~Giacalone, J.~Noronha-Hostler, M.~Luzum and J.~Y.~Ollitrault,
  arXiv:1711.08499 [nucl-th].
  
\bibitem{Novak:2013bqa} 
  J.~Novak, K.~Novak, S.~Pratt, J.~Vredevoogd, C.~Coleman-Smith and R.~Wolpert,
  Phys.\ Rev.\ C {\bf 89}, no. 3, 034917 (2014)
  [arXiv:1303.5769 [nucl-th]].
  
\bibitem{Pratt:2015zsa} 
  S.~Pratt, E.~Sangaline, P.~Sorensen and H.~Wang,
  Phys.\ Rev.\ Lett.\  {\bf 114}, 202301 (2015)
  [arXiv:1501.04042 [nucl-th]].
  
\bibitem{Bernhard:2016tnd} 
  J.~E.~Bernhard, J.~S.~Moreland, S.~A.~Bass, J.~Liu and U.~Heinz,
  Phys.\ Rev.\ C {\bf 94}, no. 2, 024907 (2016)
  [arXiv:1605.03954 [nucl-th]].


\bibitem{Bass:2017zyn} 
  S.~A.~Bass, J.~E.~Bernhard and J.~S.~Moreland,
  Nucl.\ Phys.\ A {\bf 967}, 67 (2017)
  [arXiv:1704.07671 [nucl-th]].


\bibitem{Bernhard:2017vql} 
  J.~E.~Bernhard, J.~S.~Moreland and S.~A.~Bass,
  Nucl.\ Phys.\ A {\bf 967}, 293 (2017)
  [arXiv:1704.04462 [nucl-th]].

\bibitem{Auvinen:2017fjw} 
  J.~Auvinen, I.~Karpenko, J.~E.~Bernhard and S.~A.~Bass,
  arXiv:1706.03666 [hep-ph].

\bibitem{Niemi:2015voa} 
  H.~Niemi, K.~J.~Eskola, R.~Paatelainen and K.~Tuominen,
  Phys.\ Rev.\ C {\bf 93}, no. 1, 014912 (2016)
  [arXiv:1511.04296 [hep-ph]].


\bibitem{Eskola:2017imo} 
  K.~J.~Eskola, H.~Niemi, R.~Paatelainen and K.~Tuominen,
  Nucl.\ Phys.\ A {\bf 967}, 313 (2017)
  [arXiv:1704.04060 [hep-ph]].


\bibitem{Kunszt:1992tn} 
  Z.~Kunszt and D.~E.~Soper,
  Phys.\ Rev.\ D {\bf 46}, 192 (1992).


\bibitem{Eskola:2000ji} 
  K.~J.~Eskola and K.~Tuominen,
  Phys.\ Lett.\ B {\bf 489}, 329 (2000)
  [hep-ph/0002008].


\bibitem{Eskola:2000my} 
  K.~J.~Eskola and K.~Tuominen,
  Phys.\ Rev.\ D {\bf 63}, 114006 (2001)
  [hep-ph/0010319].


\bibitem{Ellis:1985er} 
  R.~K.~Ellis and J.~C.~Sexton,
  Nucl.\ Phys.\ B {\bf 269}, 445 (1986).

\bibitem{Paatelainen:2014fsa}
  R.~Paatelainen, PhD thesis, Research Report 8/2014, University of Jy\"askyl\"a,  
  arXiv:1409.3508 [hep-ph].

\bibitem{Pumplin:2002vw} 
  J.~Pumplin, D.~R.~Stump, J.~Huston, H.~L.~Lai, P.~M.~Nadolsky and W.~K.~Tung,
  JHEP {\bf 0207}, 012 (2002)
  [hep-ph/0201195].


\bibitem{Helenius:2012wd} 
  I.~Helenius, K.~J.~Eskola, H.~Honkanen and C.~A.~Salgado,
  JHEP {\bf 1207}, 073 (2012)
  [arXiv:1205.5359 [hep-ph]].


\bibitem{Eskola:2001rx} 
  K.~J.~Eskola, K.~Kajantie and K.~Tuominen,
  Nucl.\ Phys.\ A {\bf 700}, 509 (2002)
  [hep-ph/0106330].


\bibitem{Cudell:2002xe} 
  J.~R.~Cudell {\it et al.} [COMPETE Collaboration],
  Phys.\ Rev.\ Lett.\  {\bf 89}, 201801 (2002)
  [hep-ph/0206172].


\bibitem{Antchev:2013iaa} 
  G.~Antchev {\it et al.} [TOTEM Collaboration],
  EPL {\bf 101}, no. 2, 21004 (2013).


\bibitem{Denicol:2012cn} 
  G.~S.~Denicol, H.~Niemi, E.~Molnar and D.~H.~Rischke,
  Phys.\ Rev.\ D {\bf 85}, 114047 (2012)
  Erratum: [Phys.\ Rev.\ D {\bf 91}, no. 3, 039902 (2015)]
  [arXiv:1202.4551 [nucl-th]].


\bibitem{Molnar:2013lta} 
  E.~Molnar, H.~Niemi, G.~S.~Denicol and D.~H.~Rischke,
  Phys.\ Rev.\ D {\bf 89}, no. 7, 074010 (2014)
  [arXiv:1308.0785 [nucl-th]].


\bibitem{Huovinen:2009yb} 
  P.~Huovinen and P.~Petreczky,
  Nucl.\ Phys.\ A {\bf 837}, 26 (2010)
  [arXiv:0912.2541 [hep-ph]].


\bibitem{Aamodt:2010cz} 
  K.~Aamodt {\it et al.} [ALICE Collaboration],
  Phys.\ Rev.\ Lett.\  {\bf 106}, 032301 (2011)
  [arXiv:1012.1657 [nucl-ex]].


\bibitem{Adam:2015ptt} 
  J.~Adam {\it et al.} [ALICE Collaboration],
  Phys.\ Rev.\ Lett.\  {\bf 116}, no. 22, 222302 (2016)
  [arXiv:1512.06104 [nucl-ex]].


\bibitem{Abelev:2008ab} 
  B.~I.~Abelev {\it et al.} [STAR Collaboration],
  Phys.\ Rev.\ C {\bf 79}, 034909 (2009)
  [arXiv:0808.2041 [nucl-ex]].


\bibitem{Adler:2004zn} 
  S.~S.~Adler {\it et al.} [PHENIX Collaboration],
  Phys.\ Rev.\ C {\bf 71}, 034908 (2005)
  Erratum: [Phys.\ Rev.\ C {\bf 71}, 049901 (2005)]
  [nucl-ex/0409015].


\bibitem{ALICE:2011ab} 
  K.~Aamodt {\it et al.} [ALICE Collaboration],
  Phys.\ Rev.\ Lett.\  {\bf 107}, 032301 (2011)
  [arXiv:1105.3865 [nucl-ex]].

\end{thebibliography}
\end{document}